\documentclass[11pt]{article}
\usepackage{a4p}
\usepackage{amssymb}
\usepackage{epsfig}
\usepackage{cite}
\usepackage{pennames}
\usepackage{rotating}

\newcommand{    \EhadrI}     {0.011}
\newcommand{   \EsqrtsI}     {0.007}
\newcommand{      \EmwI}     {0.004}
\newcommand{    \EfragI}     {0.007}
\newcommand{  \EbgxsecI}     {0.006}
\newcommand{  \EbgcompI}     {0.010}
\newcommand{   \EcfracI}     {0.006}
\newcommand{ \EudscompI}     {0.007}
\newcommand{ \EvtxrecoI}     {0.016}
\newcommand{   \EclifeI}     {0.003}
\newcommand{   \EcmultI}     {0.010}
\newcommand{   \ElepidI}     {0.012}
\newcommand{  \ElepmomI}     {0.003}
\newcommand{  \EbrctolI}     {0.006}

\newcommand{    \EhadrII}     {0.012}
\newcommand{   \EsqrtsII}     {0.005}
\newcommand{      \EmwII}     {0.003}
\newcommand{    \EfragII}     {0.007}
\newcommand{  \EbgxsecII}     {0.005}
\newcommand{  \EbgcompII}     {0.009}
\newcommand{   \EcfracII}     {0.007}
\newcommand{ \EudscompII}     {0.005}
\newcommand{ \EvtxrecoII}     {0.017}
\newcommand{   \EclifeII}     {0.002}
\newcommand{   \EcmultII}     {0.010}
\newcommand{   \ElepidII}     {0.014}
\newcommand{  \ElepmomII}     {0.003}
\newcommand{  \EbrctolII}     {0.006}

\newcommand{   \RcvalueI}    {0.493}
\newcommand{    \RcstatI}    {0.090}
\newcommand{    \RcsystI}    {0.032}

\newcommand{   \RcvalueII}    {0.478}
\newcommand{    \RcstatII}    {0.047}
\newcommand{    \RcsystII}    {0.032}

\newcommand{\Rcvalue}   {0.481}
\newcommand{ \Rcstat}   {0.042}
\newcommand{ \Rcsyst}   {0.032} %% updated 8.2.00

\newcommand{\Rcfinal} {\Rcvalue \pm \Rcstat \stat \pm \Rcsyst \syst}

\newcommand{  \Vcsvalue}    {0.93} %% updated 8.2.00
\newcommand{   \Vcsstat}    {0.08} %% updated 8.2.00
\newcommand{   \Vcssyst}    {0.06} %% updated 8.2.00
\newcommand{    \Vcsckm}    {0.004} %% updated 8.2.00
\newcommand{    \Vcserr}    {0.10} %% updated 8.2.00
\newcommand{\Vcsfinal} {\Vcsvalue \pm \Vcsstat \stat \pm \Vcssyst \syst}
\newcommand{\Vcssummary} {\Vcsvalue \pm \Vcserr}

\newcommand{  \GammaW}    {698} %% updated 19.7.00
\newcommand{ \Gamstat}     {61} %% updated 19.7.00
\newcommand{ \Gamsyst}     {46} %% updated 19.7.00
\newcommand{ \Gambrww}     {18} %% updated 19.7.00

\newcommand{\GammaWfinal} {[\GammaW \pm \Gamstat \stat \pm \Gamsyst \syst \pm \Gambrww \wwext]\,\MeV}

\newcommand{  \SumVij}    {0.990} %% updated 19.7.00
\newcommand{ \Sumstat}    {0.087} %% updated 19.7.00
\newcommand{ \Sumsyst}    {0.065} %% updated 19.7.00
\newcommand{ \Sumbrww}    {0.026} %% updated 19.7.00
\newcommand{\SumVijfinal} {\SumVij \pm \Sumstat \stat \pm \Sumsyst \syst \pm \Sumbrww \wwext}

\newcommand{  \VcsvalueG}    {0.969} %% updated 19.7.00
\newcommand{   \VcsstatG}    {0.045} %% updated 19.7.00
\newcommand{   \VcssystG}    {0.034} %% updated 19.7.00
\newcommand{   \VcsbrwwG}    {0.013} %% updated 19.7.00
\newcommand{    \VcsckmG}    {0.004} %% updated 19.7.00
\newcommand{    \VcserrG}    {0.058} %% updated 19.7.00

\newcommand{\VcsfinalG} {\VcsvalueG \pm \VcsstatG \stat \pm \VcssystG \syst \pm \VcsbrwwG \wwext \pm \VcsckmG \ckmerr}
\newcommand{\VcssummaryG} {\VcsvalueG \pm \VcserrG}

\newcommand{\uquark}{\ensuremath{\mathrm{u}}}
\newcommand{\dquark}{\ensuremath{\mathrm{d}}}

\newcommand{\squark}{\ensuremath{\mathrm{s}}}

\newcommand{\WW}{\ensuremath{\mathrm{WW}}}
\newcommand{\NWqqlvI}{350}
\newcommand{\NWqqqqI}{432}
\newcommand{\NWqqlvII}{1235}
\newcommand{\NWqqqqII}{1532}

\newcommand{\lumiI}{56.5}
\newcommand{\lumiII}{180.2}
\newcommand{\rrootsI}{182.7}
\newcommand{\rrootsII}{188.6}

\newcommand{\epem}{\ensuremath{\mathrm{e}^+\mathrm{e}^-}}

\newcommand{\lplm}{\ensuremath{\ell^+\ell^-}}
\newcommand{\Zz}{\ensuremath{{\mathrm{Z}^0}}}

\newcommand{\WpWm}{\ensuremath{\mathrm{W}^+\mathrm{W}^-}}
\newcommand{\Wboson}{\ensuremath{\mathrm{W}}}
\newcommand{\Zboson}{\ensuremath{\mathrm{Z}}}

\newcommand{\eeWW}{\ensuremath{\epem\rightarrow\WpWm}}
\newcommand{\qq}{\ensuremath{\mathrm{q\overline{q}}}}
\newcommand{\bb}{\ensuremath{\mathrm{b\overline{b}}}}

\newcommand{\lnu}{\ensuremath{\ell\overline{\nu}_{\ell}}}
\newcommand{\lv}{\lnu}
\newcommand{\lpnu}{\ensuremath{\ell^+ \nu_{\ell}}}
\newcommand{\lmnu}{\ensuremath{{\ell^{\prime}}^-\overline{\nu}_{\ell^{\prime}}}}
\newcommand{\enu}{\ensuremath{\mathrm{e\overline{\nu}_{e}}}}
\newcommand{\mnu}{\ensuremath{\mu\overline{\nu}_{\mu}}}
\newcommand{\tnu}{\ensuremath{\tau\overline{\nu}_{\tau}}}

\newcommand{\qqqq}{\ensuremath{\qq\qq}}

\newcommand{\qqln}{\ensuremath{\qq\lnu}}
\newcommand{\qqll}{\ensuremath{\qq\lplm}}

\newcommand{\WWqqln}{\ensuremath{\WpWm\rightarrow\qq\lnu}}
\newcommand{\WWqqlv}{\WWqqln}

\newcommand{\WWqqqq}{\ensuremath{\WpWm\rightarrow\qq\qq}}

\newcommand{\Wcx}{\ensuremath{\W\to\rm c\,\rm X}}
\newcommand{\Wux}{\ensuremath{\W\to\rm light~\rm flavour}}

\newcommand{\WWlvlv}{\ensuremath{\WpWm\rightarrow\lpnu\lmnu}}

\newcommand{\Zqq}{\ensuremath{\Zz/\gamma\rightarrow\qq}}
\newcommand{\Zbb}{\ensuremath{\Zz/\gamma\rightarrow\bb}}

\newcommand{\Mw}{\ensuremath{M_{\mathrm{W}}}}

\newcommand{\Jetset}{\mbox{J{\sc etset}}}
\newcommand{\Koralw}{\mbox{K{\sc oralw}}}

\newcommand{\Excalibur}{\mbox{E{\sc xcalibur}}}

\newcommand{\grcff}{{\scshape grc4f}}
\newcommand{\Pythia}{{\scshape Pythia}}

\newcommand{\Herwig}{{\scshape Herwig}}

\newcommand{\half}{\ensuremath{\textstyle\frac{1}{2}}}

\newcommand{\GeV}{\ensuremath{\mathrm{GeV}}}
\newcommand{\pb}{\ensuremath{\mathrm{pb}}}

\newcommand{\Wlv}{\mbox{$\mathrm{W}\rightarrow\lnu$}}
\newcommand{\Wqq}{\mbox{$\mathrm{W}\rightarrow\qq$}}

\newcommand{\qqen}{\ensuremath{\qq\enu}}
\newcommand{\qqev}{\qqen}

\newcommand{\qqmn}{\ensuremath{\qq\mnu}}
\newcommand{\qqmv}{\qqmn}
\newcommand{\qqtn}{\mbox{\qq\tnu}}
\newcommand{\qqtv}{\qqtn}

\newcommand{\Vud} {\ensuremath{|V_{\mathrm{ud}}|}}
\newcommand{\Vus} {\ensuremath{|V_{\mathrm{us}}|}}
\newcommand{\Vcd} {\ensuremath{|V_{\mathrm{cd}}|}}
\newcommand{\Vcb} {\ensuremath{|V_{\mathrm{cb}}|}}
\newcommand{\Vub} {\ensuremath{|V_{\mathrm{ub}}|}}
\newcommand{\Vcs} {\ensuremath{|V_{\mathrm{cs}}|}}
\newcommand{\mW}{\ensuremath{M_{\mathrm{W}}}}

\newcommand{\sub}[1]{\ensuremath{_{\mathrm{#1}}}}

\newcommand{\stat}{\ensuremath{{\mathrm{\,(stat.)}}}}
\newcommand{\syst}{\ensuremath{{\mathrm{\,(syst.)}}}}
\newcommand{\wwext}{\ensuremath{{\mathrm{\,(ext.)}}}}
\newcommand{\ckmerr}{\ensuremath{{\mathrm{\,(CKM)}}}}
\newcommand{\der}{\ensuremath{{\mathrm{d}}}}

\def\etal{\mbox{{\it et al.}}}
\def\ie{\mbox{{\it i.e.}}}

\def\gappeq{\ensuremath{\mathrel{ \rlap{\raise.5ex\hbox{>}}
                      {\lower.5ex\hbox{\sim}}}}}
\def\lappeq{\ensuremath{\mathrel{ \rlap{\raise.5ex\hbox{<}}
                      {\lower.5ex\hbox{\sim}}}}}
\newcommand{\PLB}[3]  {Phys.\ Lett.\ \textbf{B#1} (#2) #3}
\newcommand{\ZPC}[3]  {Z.\ Phys.\ \textbf{C#1} (#2) #3}
\newcommand{\EPC}[3]  {Eur.\ Phys.\ J.\ \textbf{C#1} (#2) #3}
\newcommand{\NIMA}[3] {Nucl.\ Instr.\ Meth.\ \textbf{A#1} (#2) #3}

\newcommand{\PRL}[3]  {Phys.\ Rev.\ Lett.\ \textbf{#1} (#2) #3}
\newcommand{\PRD}[3]  {Phys.\ Rev.\ \textbf{D#1} (#2) #3}
\newcommand{\NPB}[3]  {Nucl.\ Phys.\ \textbf{B#1} (#2) #3}
\newcommand{\CiP}[3]  {Comp.\ in Phys.\ \textbf{#1} (#2) #3}

\newcommand{\CPC}[3]  {Comput.\ Phys.\ Commun.\ \textbf{#1} (#2) #3}
\newcommand{\JPH}[3] {J.~Phys.\ \textbf{#1} (#2) #3}
\def\opalackerstaff{OPAL Collaboration, K.\ Ackerstaff \etal}
\def\opalabbiendi{OPAL Collaboration, G. Abbiendi \etal}
\def\opalalexander{OPAL Collaboration, G.\ Alexander \etal}

\newcommand{\Br}{\ensuremath{\mathrm{Br}}}
\newcommand {\Rc}{\ensuremath{R_{\mathrm{c}}^{\mathrm{W}}}}

\newcommand {\W}{\ensuremath{{\mathrm{W}}}}

\newcommand {\uds}{\ensuremath{\mathrm{uds}}}

\newcommand {\epsu}{\ensuremath{\epsilon_{\mathrm{W \rightarrow u}}}}
\newcommand {\epsd}{\ensuremath{\epsilon_{\mathrm{W \rightarrow d}}}}
\newcommand {\epss}{\ensuremath{\epsilon_{\mathrm{W \rightarrow s}}}}
\newcommand {\epsc}{\ensuremath{\epsilon_{\mathrm{W \rightarrow c}}}}

\newcommand {\epsbgd}{\ensuremath{\epsilon_{\mathrm{bgd}}}}

\newcommand {\qqlv}{\ensuremath{\mathrm{q\overline{q}\ell\overline{\nu}}}}
\newcommand{\GF}{\ensuremath{G_{\mathrm{F}}}}
\newcommand{\alphas} {\ensuremath {\alpha_{\mathrm{s}}}}
\newcommand{\Wtohad} {\ensuremath{\mathrm{W \rightarrow hadrons}}}
\newcommand{\BRctol} {\ensuremath{\Br(\mathrm{c \rightarrow \ell})}}
\newcommand{\GammatotW} {\ensuremath {\Gamma_{\mathrm{tot}}^{\mathrm{W}}}}

\newcommand{\BigR} {\ensuremath {{\cal{R}}}}

\newcommand{\Wtoev} {\ensuremath{\mathrm{W \rightarrow e \bar{\nu}}}}

\newcommand{\Ztoee} {\ensuremath{\mathrm{Z^0 \rightarrow e^+ e^-}}}
\newcommand{\ppbar} {\ensuremath {{\mathrm{p \overline{p}}}}}

\newcommand{\MeV}{\ensuremath{\mathrm{MeV}}}

\newcommand {\um}{\ensuremath{\mathrm{\mu m}}}

\begin{document}
\begin{titlepage}
\begin{center}{\large   EUROPEAN ORGANIZATION FOR NUCLEAR RESEARCH
}\end{center}\bigskip
\begin{flushright}
%       OPAL PR322 \\
       CERN-EP-2000-100   \\ 
       July $20^{\rm th}$ 2000
\end{flushright}
\bigskip\bigskip\bigskip\bigskip\bigskip
\begin{center}{\huge\bf   A Measurement of the Rate of Charm Production in \boldmath \W\ Decays
}\end{center}\bigskip\bigskip
\begin{center}{\LARGE The OPAL Collaboration
}\end{center}\bigskip\bigskip
\bigskip\begin{center}{\large  Abstract}\end{center}
    Using data recorded at centre--of--mass energies around
    183\,GeV and 189\,GeV with the OPAL detector at LEP, the fundamental
    coupling of the charm quark 
    to the \Wboson\ boson has been studied.  The ratio
    $\Rc \equiv \Gamma\,(\W \rightarrow \mathrm{c\,X}) / 
    \Gamma\,(\W \rightarrow \mathrm{hadrons})$
    has been measured from jet properties, lifetime 
    information, and leptons produced in charm decays.
    A value compatible with the Standard Model expectation of 0.5 
    is obtained: $\Rc = \Rcfinal$. 
    By combining this result with measurements of the \W\ boson 
    total width and hadronic branching ratio, the magnitude of 
    the CKM matrix element \Vcs\ is determined to be
    $\Vcs = \VcssummaryG$.
\bigskip\bigskip\bigskip\bigskip
\bigskip\bigskip
\begin{center}{\large
(Submitted to Physics Letters B)
}\end{center}
\end{titlepage}
\begin{center}{\Large        The OPAL Collaboration
}\end{center}\bigskip
\begin{center}{
%begin authorlist PLEASE DO NOT DELETE THIS COMMENT
G.\thinspace Abbiendi$^{  2}$,
K.\thinspace Ackerstaff$^{  8}$,
C.\thinspace Ainsley$^{  5}$,
P.F.\thinspace {\AA}kesson$^{  3}$,
G.\thinspace Alexander$^{ 22}$,
J.\thinspace Allison$^{ 16}$,
K.J.\thinspace Anderson$^{  9}$,
S.\thinspace Arcelli$^{ 17}$,
S.\thinspace Asai$^{ 23}$,
S.F.\thinspace Ashby$^{  1}$
% ...
%end authorlist PLEASE DO NOT DELETE THIS COMMENT
}\end{center}\bigskip
\bigskip
%begin institutes
$^{  1}$School of Physics and Astronomy, University of Birmingham,
Birmingham B15 2TT, UK
\newline
$^{  2}$Dipartimento di Fisica dell' Universit\`a di Bologna and INFN,
I-40126 Bologna, Italy
\newline
$^{  3}$Physikalisches Institut, Universit\"at Bonn,
D-53115 Bonn, Germany
\newline
% ...
$^{  5}$Cavendish Laboratory, Cambridge CB3 0HE, UK
\newline
% ...
$^{  8}$CERN, European Organisation for Nuclear Research,
CH-1211 Geneva 23, Switzerland
\newline
$^{  9}$Enrico Fermi Institute and Department of Physics,
University of Chicago, Chicago IL 60637, USA
\newline
% ...
$^{ 16}$Department of Physics, Schuster Laboratory, The University,
Manchester M13 9PL, UK
\newline
$^{ 17}$Department of Physics, University of Maryland,
College Park, MD 20742, USA
\newline
% ...
$^{ 22}$Department of Physics and Astronomy, Tel Aviv University,
Tel Aviv 69978, Israel
\newline
$^{ 23}$International Centre for Elementary Particle Physics and
Department of Physics, University of Tokyo, Tokyo 113-0033, and
Kobe University, Kobe 657-8501, Japan
\newline
% ...
$^{  c}$ and Institute of Nuclear Research, Debrecen, Hungary
\newline
% ...
\newpage

\section{Introduction}
\label{intro} 

Since 1997 the LEP \epem\ collider has been operated at 
energies above the threshold for W-pair production.
This offers a
unique opportunity to study the hadronic decays of \Wboson\ bosons in a clean
environment and to investigate the coupling strength of \Wboson\ bosons
to different quark flavours. The fraction of \Wboson\ bosons 
decaying hadronically to different quark flavours
is proportional to the sum of the 
squared magnitudes of the corresponding elements of the
Cabibbo--Kobayashi--Maskawa (CKM) matrix~\cite{bib:CKMmatrix}. 
A measurement of 
the production rates of different flavours therefore
gives access to the individual CKM matrix elements. 

In \eeWW\ events, the only heavy 
quark commonly produced is the charm quark.
The production of bottom quarks is in fact highly suppressed
due to the small magnitude of \Vub\ and \Vcb\ and
the large mass of the top quark, the weak partner of the bottom quark.
This allows for
a direct measurement of the production fraction 
of charm in W decays
without a separation of charm and bottom quarks.
The magnitude of the CKM matrix element $V_{\mathrm{cs}}$ can 
then be derived 
from the charm production rate, using the knowledge of the 
other CKM matrix elements. So far direct measurements of \Vcs\ have 
limited precision compared to the other CKM matrix elements important 
in \Wboson\ decays, \ie\ \Vud, \Vus, and \Vcd.
The most recent evaluation of the magnitude of $V_{\mathrm{cs}}$ from
exclusive charm semielectronic decays
yields $\Vcs = 1.04 \pm 0.16$~\cite{bib:PDG}.

In the analysis presented here, a value of the partial decay
width $\Rc \equiv \Gamma\,(\W \rightarrow \mathrm{c\,X}) / 
\Gamma\,(\W \rightarrow \mathrm{hadrons})$
is extracted from the properties of final state particles in \W\ decays. 
Charm hadron identification is
based upon jet properties, lifetime information,
and semileptonic decay products in \WWqqqq\ and \WWqqlv\ events.
The measured value of \Rc\ is then used to determine \Vcs. 

%%%%%%%%%%%%%%%%%%%%%%%%%%%% Monte Carlo samples %%%%%%%%%%%%%%%%%%%%%%%%

\section{Data and Monte Carlo Samples}
\label{montecarlo} 

The data used for this analysis were recorded at LEP
by the OPAL detector~\cite{bib:OPALdetector} at
centre--of--mass energies around 183\,GeV in 1997 and 189\,GeV in
1998.  OPAL\footnote{The OPAL
coordinate system is defined such that the $z$-axis is parallel to and in the 
direction of the $e^{-}$ beam, the $x$-axis lies in the plane of the 
accelerator pointing towards the centre of the LEP ring, and the $y$-axis is
normal to the plane of the accelerator and has its positive direction defined 
to yield a right-handed coordinate system.  The azimuthal angle, $\phi$, and 
the polar angle, $\theta$, are the conventional spherical coordinates.} 
is a multipurpose high 
energy physics detector incorporating excellent charged and neutral 
particle detection and measurement capabilities.
The integrated luminosities used for the analysis 
amount to $\lumiI\,\pb^{-1}$\,($\lumiII\,\pb^{-1}$) at 
luminosity--weighted mean centre--of--mass energies 
of \rrootsI \,GeV\,(\rrootsII \,GeV).
In addition, $2.1\,\pb^{-1}$ ($3.1\,\pb^{-1}$) of calibration 
data were collected at $\sqrt{s} \sim M_{\Zboson}$ in 1997 (1998)
and have been used for fine tuning of the Monte Carlo simulation.

To determine the selection criteria and tagging efficiencies,
Monte Carlo samples were generated using the 
\Koralw\,1.42~\cite{bib:KORALW} program, 
which utilizes \Jetset\,7.4~\cite{bib:JETSET} for fragmentation,
followed by a full simulation of the OPAL detector~\cite{bib:GOPAL}.
The centre--of--mass energies were set to
$\sqrt{s}=183$\,GeV and 189\,GeV with a \Wboson\ mass of $\mW = 80.33$\,GeV. 
Additional samples with different centre--of--mass energies and
\Wboson\ masses were used to estimate the sensitivity to these 
parameters. 
Further samples generated with \Pythia~\cite{bib:PYTHIA},
\Herwig~\cite{bib:HERWIG}, and \Excalibur~\cite{bib:Excalibur}
were used to determine the model dependence of the measurement.

The dominant background sources for this analysis are \Zqq\
events and four--fermion final states which are not from 
\WpWm\ decays. 
Background samples were generated using the \Pythia, \Herwig, 
\Excalibur, and \grcff~\cite{bib:GRC4F} generators. More details 
on the treatment 
of the backgrounds can be found in References~\cite{bib:PR260, bib:PN378}.

%%%%%%%%%%%%%%%%%%%%%%%%%%%%%%% Event selection %%%%%%%%%%%%%%%%%%%%%%%%

\section{Event Selection}
\label{selection} 

The \WW\ event selections are the same as those used for the OPAL \WW\ production 
cross--section measurements
at $\sqrt{s} = 183$\,GeV~\cite{bib:PR260} and 189\,GeV~\cite{bib:PN378}.
Events not selected as \WWlvlv\ candidates are passed to
the \WWqqlv\ selection procedure and may be classified as 
\qqev, \qqmv, or \qqtv\  candidates.
The \WWqqqq\ selection is applied to events that fail the preceding
selections. It has been checked that the event selection introduces a 
negligible bias in the flavour composition. In this analysis, the small 
fraction of selected singly produced W bosons ($\mathrm{W\enu}$) which
decay hadronically to a charm quark is considered as signal.

Since the identification of charm mesons relies heavily on 
lifetime and lepton information, the relevant parts of the 
detector were required to be operational while the data were
recorded. Thus, in 
addition to the requirements which were already applied 
in References~\cite{bib:PR260, bib:PN378}, it was required that 
the silicon microvertex detector and the muon chambers were fully operational. 
After all cuts, a total of \NWqqlvI\ (\NWqqlvII) \WWqqlv\ candidates and  
\NWqqqqI\ (\NWqqqqII) \WWqqqq\ candidates were selected 
at 183\,GeV (189\,GeV). A breakdown  of the different background 
contributions is given in Table~\ref{tab:NW189}.
The expected number of background events is calculated from the
background cross--sections given in References~\cite{bib:PR260,
bib:PN378}; the errors correspond to the systematic errors quoted in
these references. 

%---------------------------TABLE NW---------------------------------
\begin{table}[t]
\begin{center}
\begin{tabular}{|l||c|c|c|c|}
\hline
                 & \multicolumn{2}{|c|}{183 GeV}       & \multicolumn{2}{|c|}{189 GeV}  \\
\hline
Type             & \qqlv           &    \qqqq         & \qqlv           &    \qqqq    \\
\hline\hline
Selected Events  &  \NWqqlvI       &  \NWqqqqI        & \NWqqlvII       & \NWqqqqII \\
\hline \hline
\Zqq             & $15.4 \pm 1.6$  & $77.4 \pm 8.5$   & $~45.5 \pm 5.6$  & $240.0 \pm 15.9$ \\
\qqqq            & $~0.5 \pm 0.6$  & $12.4 \pm 2.8$   &$\,~~2.5 \pm 0.6$ & $~70.3 \pm 12.3$ \\
\qqll            & $10.2 \pm 1.2$  & $~4.1 \pm 0.3$   & $~27.6 \pm 2.7$  & $~~8.6 \pm ~1.4$ \\
$\mathrm{W\enu}$ & $~6.2 \pm 1.7$  & $~~0 \pm 1$      & $~23.3 \pm 4.1$  & $~~0 \pm 1$ \\
\hline
Total Background & $32.4 \pm 2.7$  & $93.9 \pm 9.0$   & $~98.9 \pm 7.5$  & $318.9\pm 20.2$ \\
\hline
\end{tabular}
\end{center}
\caption{
\label{tab:NW189}
Number of selected \WW\ candidate events and those 
expected from the main background sources for the 
selection at 183\,GeV and 189\,GeV.
The ``$\qqqq$'' background source refers to processes
excluding doubly resonant \W-pair production at tree level 
(CC03 diagrams).
The entry labeled  ``$\mathrm{W\enu}$'' refers to events where a singly 
produced \Wboson\ decays hadronically.
The errors are the systematic errors on the background contributions,
as evaluated in References~\cite{bib:PR260,bib:PN378}.
}
\end{table}

%%%%%%%%%%%%%%%%%%%%%%%%%%%%%%% Charm tagging %%%%%%%%%%%%%%%%%%%%%%%%%%

\section{\boldmath Measurement of \Rc}
\label{charmtag} 

After the event selection, the hadronic part (which excludes the identified lepton)
of a \qqlv\ event is forced into two jets
using the Durham algorithm~\cite{bib:Durham}, while \qqqq\ events are forced 
into four jets.  Subsequently, a relative likelihood discriminant 
is calculated for each jet to separate jets 
originating from charm quarks and jets from u, d, and s quarks. 
This discriminant relies on jet and event shape properties,
lifetime information, and lepton identification as described in the 
remaining parts of this section.
The resulting likelihood distribution 
is fitted to obtain the relative fractions of \uds\ and
charm jets. 

In \qqln\ events the two jets can be uniquely assigned to the
decay of one \Wboson\ boson, while
for \qqqq\ events the four jets have to be grouped into pairs 
that are
assumed to originate from the decay of one \Wboson\ boson.
The jet pairing is chosen using a kinematic fit to find the 
jet combination which is most compatible with the production of 
two on-shell \Wboson\ bosons with a mass of 80.33\,GeV.
After the jet pairing has been performed in \qqln\ and \qqqq\ events, 
the jet energies are calculated in a 
kinematic fit, imposing energy--momentum
conservation and equality of the masses of both \Wboson\ candidates.
If the kinematic fit fails, the jet energies 
of each jet pair are scaled so that their
sum equals the beam energy. The choice of the jet pairing 
influences the result of the present analysis only 
through the values of the jet angle $\theta_{\mathrm{jet}}$ 
in the di-jet rest frame and the di-jet angle $\theta_{\mathrm{W}}$
in the laboratory frame, both of which are calculated from the 
jet momenta determined by the kinematic fit.
These two angles are employed in the likelihood functions
as described in Sections~\ref{leptontag} and~\ref{combinedtag}.
It has been checked in the \Wboson\ mass measurement 
analysis~\cite{bib:Wmass183,bib:Wmass189}
and in an analysis concerned with properties of hadronically decaying 
\Wboson\ bosons~\cite{bib:PN342} that jet properties relevant for the jet
pairing algorithm and the kinematic fit are well modelled in the 
Monte Carlo simulation.

%%%%%%%%%%%%%%%%%%%%%%%%%%%%%%% Lifetime tag %%%%%%%%%%%%%%%%%%%%%%%%%%%

\subsection{Secondary Vertex Tag}
\label{lifetimetag} 

Weakly decaying charm hadrons have lifetimes between 
0.2\,ps and 1\,ps~\cite{bib:PDG},
leading to typical decay lengths of a few hundred microns to
a few millimetres at LEP2 energies.
These relatively long--lived particles produce secondary decay
vertices which are significantly displaced from the primary 
event vertex. 
The primary vertex in an event is found by fitting all 
charged tracks to a common point in space~\cite{bib:teardown}. 
Its determination is 
improved by using the average beam position as a constraint
with RMS values of
$150\,\um$ in $x$, dominated by the width of 
the beam, about $20\,\um$ in $y$, and less than 1\,cm in $z$. 

In the search for charm decay products, tracks in the central 
detector and clusters in the electromagnetic calorimeter within a given 
jet are considered. 
Tracks used to find secondary vertices must fulfill the quality criteria
used in the \WW\ cross-section analyses~\cite{bib:PR260,bib:PN378}.
Additionally, the tracks are required to have a momentum
larger than $0.75$\,GeV, a signed distance of closest
approach to the primary vertex in the $r\phi$
plane (2D-impact parameter) 0\,cm $< d_0<$ 0.3\,cm,
and an error on the impact parameter, $\sigma_{d_0}$, smaller
than 0.1\,cm.
From the tracks which fulfill these criteria,
the tracks which are most likely to originate
from the decay products of a charm hadron are identified via 
an iterative procedure. 
Particles from charm hadron decays tend to have a large momentum
and a large rapidity
$y = \frac{1}{2} \ln \left ( \frac{E+ p_\parallel}{E-p_\parallel} \right )$
(where $p_\parallel$ is the particle's momentum component parallel to the jet 
direction and $E$ is its energy) compared to fragmentation tracks, and
their combined invariant mass should be compatible with a charm hadron mass. 
Therefore for each track and
electromagnetic cluster which is not associated to any track,
the rapidity $y$ is calculated assuming the charged
pion mass for tracks and zero mass for unassociated clusters.
The track or cluster with the lowest rapidity is
removed from the sample and the invariant mass of the
remaining tracks and clusters is calculated.
This procedure is repeated until
the mass of the remaining group of tracks and clusters falls below
$2.5$\,GeV. This procedure selects preferentially tracks
from the decay of charm hadrons~\cite{bib:OPALRc}.
 
All selected tracks in the jet are fitted to a common vertex 
in three dimensions. 
Tracks which contribute more than 4 to the $\chi^2$ of this 
fit are removed and the fit is repeated. The procedure is 
stopped when all tracks pass the $\chi^2$ requirement.
At least two tracks are required to remain in the fit for the
secondary vertex finding to be successful.
Once tracks are associated to a secondary vertex, the distance $L$ between 
the primary and the secondary vertex and its error $\sigma_L$
are computed. 
The decay length 
significance of a secondary vertex, $L/\sigma_L$, is required to be 
positive ({\it{i.e.}} $L/\sigma_L>0$), where $L$ is given a positive 
sign if the secondary vertex is displaced from the primary vertex along 
the direction of the jet momentum.

Jets containing a good secondary vertex are then used
to tag charm decay products by means of an 
artificial neural network (ANN)~\cite{bib:Jetnet} 
to enhance the charm purity at a given efficiency. The nine ANN input 
variables are: the decay length 
significance $L/\sigma_L$ of the secondary vertex;
the primary vertex joint probability (for a definition, see
Reference~\cite{bib:joint});
$E_{\rm vertex}/E_{\rm beam}$, where $E_{\rm vertex}$ is the 
sum of the energy of all the tracks assigned to the
secondary vertex and $E_{\rm beam}$ is the beam energy;
the distance of closest approach to the primary event vertex 
in $r \phi z$ (3D) of the pseudo--particle formed  from all the tracks 
in the secondary vertex; the normalised weighted average (weighted by 
the measured error) of all the track 2D-impact parameters within 
the secondary vertex; the highest momentum of the charged particles 
in the secondary vertex; the invariant mass of the vertex; the largest 
track 3D-impact parameter in the secondary vertex; and the second 
largest track 3D-impact parameter in the secondary vertex.

\begin{figure}
\unitlength1cm
  \epsfxsize=\textwidth
  \epsfbox{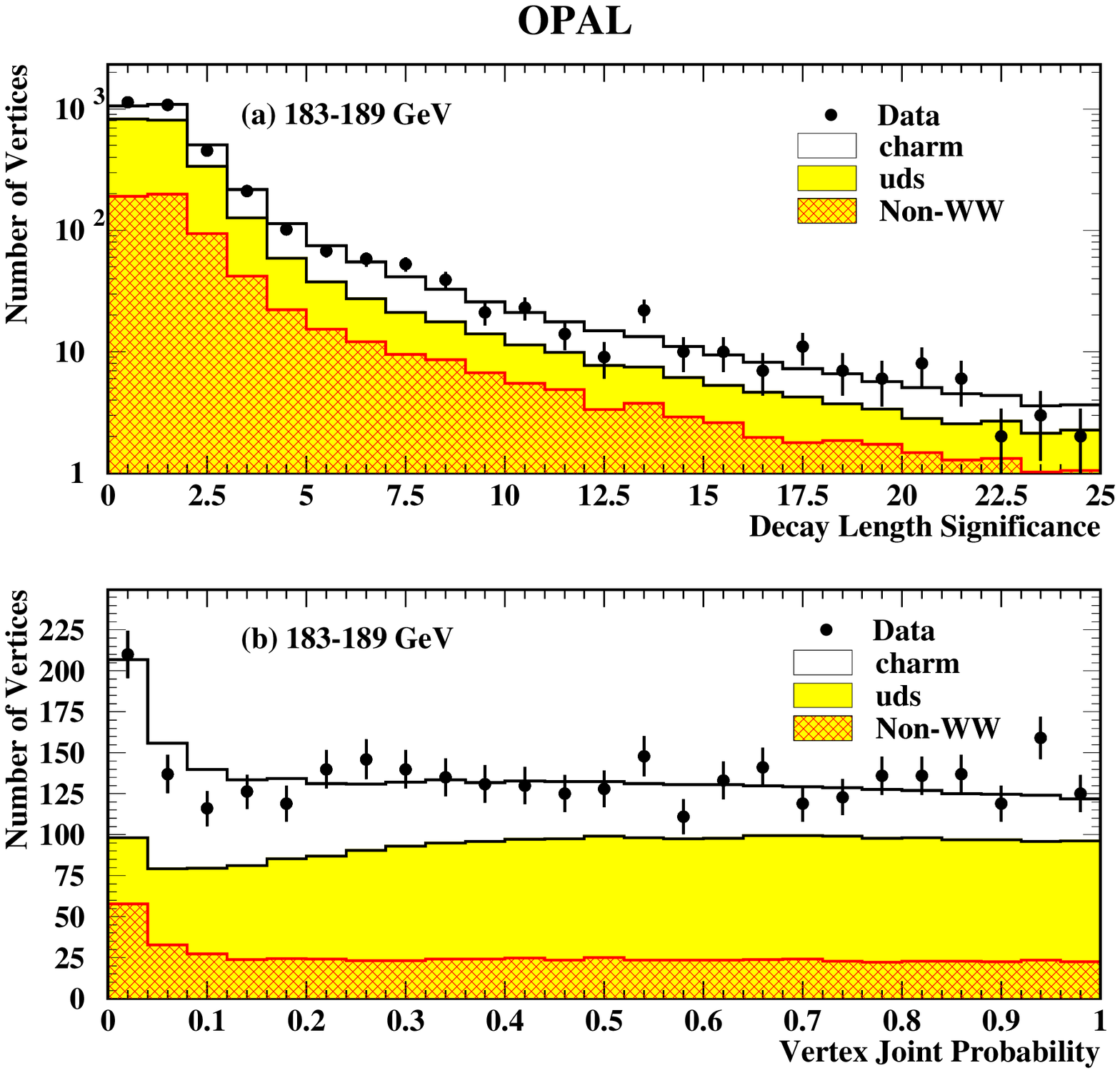}
\caption[]
        {\label{fig:vtxcontrol}
        Distributions of the most sensitive input variables
        of the vertex ANN at $\sqrt{s}=183-189$\,GeV:
        (a) decay length significance $L/\sigma_L$ and (b) primary vertex 
        joint probability.
        Points with error bars represent the data (183\,GeV and 189\,GeV samples 
        combined). Histograms represent the Monte Carlo expectation.}
\end{figure}

\begin{figure}
\unitlength1cm
  \epsfxsize=\textwidth
  \epsfbox{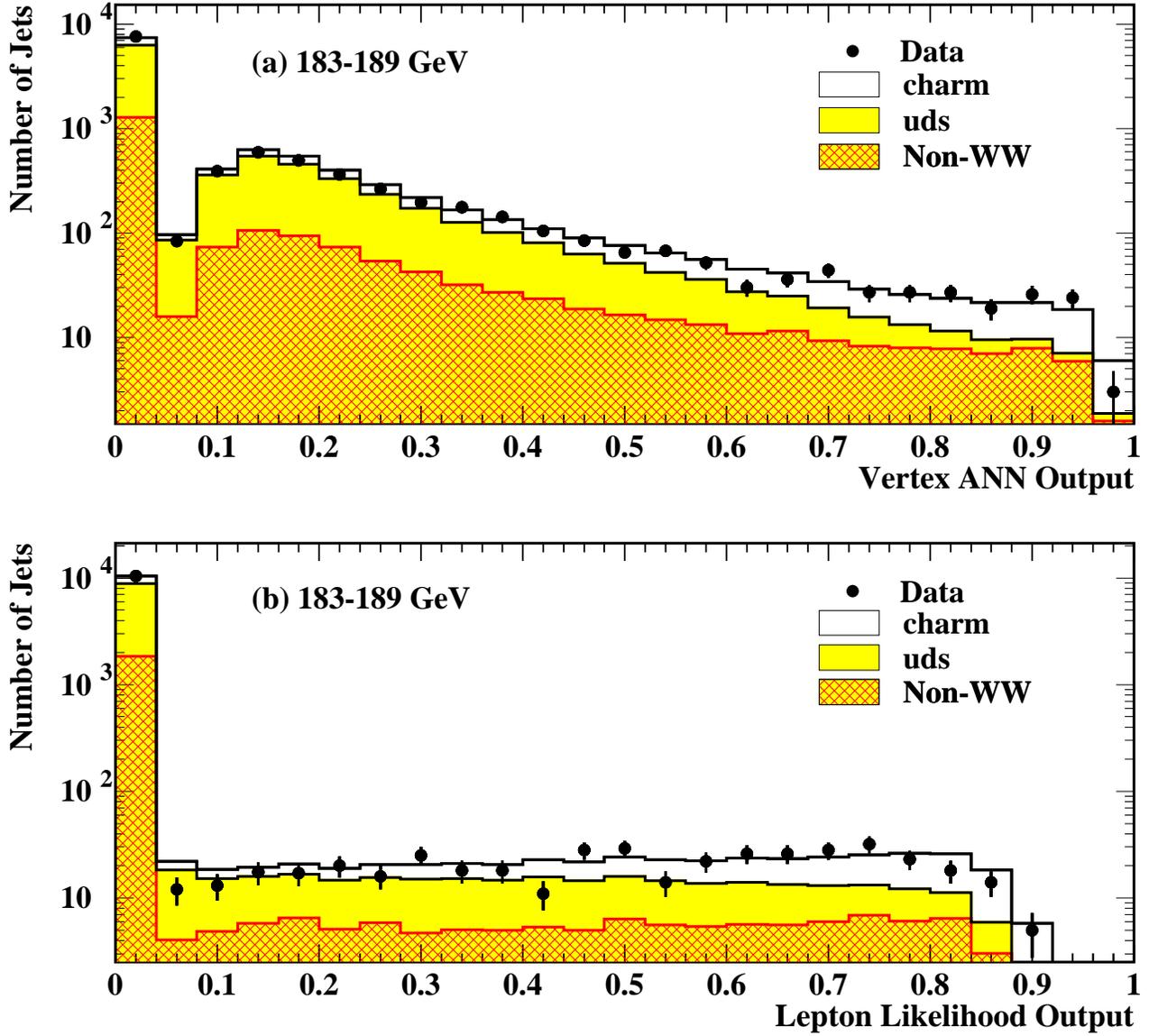}
\caption[]
        {\label{fig:tagcontrol}
        Distributions of (a) the vertex ANN output and (b) the likelihood 
        output of the lepton tag
        at $\sqrt{s}=183-189$\,GeV. Points with error bars represent the 
        data (183\,GeV and 189\,GeV samples combined). 
        Histograms show the results of the fits to the combined 
        likelihood described in Section~\ref{Rcfit}. If no vertex or lepton tag 
        information is available, the corresponding variable is assigned 
        the value of zero.}
\end{figure}

The same vertex ANN is used for \WWqqqq\ and \WWqqln\ events.
The distributions of the two most discriminating input variables
for the vertex ANN 
are shown in Figure~\ref{fig:vtxcontrol}. Typically, charm
hadrons have secondary vertices which are displaced
from the primary vertex and charged tracks associated with 
charm particles have a lower probability to come from the primary 
vertex. Figure~\ref{fig:tagcontrol}a shows a comparison of the ANN output
between data and Monte Carlo for $\sqrt{s}= 183-189$\,GeV. 
The simulation of the OPAL tracking detectors~\cite{bib:GOPAL} 
requires a detailed understanding of the track-finding 
efficiency and resolution. This is achieved by tuning
the Monte Carlo simulation
with the calibration data taken at the \Zz\ resonance.
Figure~\ref{fig:syst}a shows the distribution of $L/\sigma_L$ 
for secondary vertices.
The effect of a 5\,\% variation of the tracking
resolution used to assess the systematic uncertainty
is also depicted. 

%%%%%%%%%%%%%%%%%%%%%%%%%%%%%%% Lepton tag %%%%%%%%%%%%%%%%%%%%%%%%%%%%%

\subsection{Lepton Tag}
\label{leptontag}  

About $20\,\%$ of all charm hadrons decay 
semileptonically and produce an
electron or a muon in the final state. 
Because of the relatively large mass and hard fragmentation of 
the charm quark compared to light quarks,
this lepton is expected to have a larger 
momentum than leptons from other sources (except the small contribution 
from semileptonic bottom decays in background events). 
Therefore, identified 
electrons and muons can be used as tags for charm hadrons in 
\Wboson\ boson decays. 

In a first step, the leptons are identified using ANN
algorithms. 
The electron ANN described in References~\cite{bib:NN5,bib:NN8}
is used. 
Its most important input variables are the specific ionization 
in the central jet chamber ($\der E/\der x$) of the electron candidate track
and the ratio of the associated energy deposited 
in the electromagnetic calorimeter to the momentum of this track ($E/p$).
Photon conversions are rejected from the sample using another dedicated 
ANN~\cite{bib:NN5,bib:NN8}. The
muon ANN~\cite{bib:muNN} is based on 
information from the tracking detectors (including $\der E/\der x$),
the hadronic calorimeter, and the muon chambers. The lepton candidates 
have to be well contained in the detector ($|\cos \theta_{\ell}|<0.9$,
where $\theta_{\ell}$ is the polar angle with respect to the 
$\mathrm{e^-}$ beam direction) and are required to have momentum 
$p_{\ell}>2$\,GeV.

The most energetic lepton candidate in each jet
is used to calculate a relative likelihood discriminant 
with the goal of separating
leptons in charm decays from leptons from other sources. 
Dedicated likelihood functions are set up for electrons and muons 
in \WWqqqq\ and \WWqqlv\ events, using
additional variables that describe properties of the lepton
within a jet and enhance 
the separation of charm jets from \uds\ jets.
The likelihood calculation is done using the method 
described in Reference~\cite{bib:PCpackage}.
The following input variables are used
both for electrons and for muons: the output of the 
electron or muon ANN; the magnitude of the cosine of the polar 
angle of the lepton candidate $|\cos \theta_{\ell}|$; and the magnitude 
of the momentum of the lepton candidate $p_{\ell}$.

In \qqqq\ events, where the charge of the decaying W boson 
is not known, the likelihood function also uses the product of the lepton 
charge and the cosine of the 
W production angle, $Q_{\ell} \,\cos \theta_\W$,
where $\theta_\W$ is defined as the polar angle of the sum of 
momenta of both jets of the pair that contains the lepton.
This quantity makes use of the strong forward--backward asymmetry of
the produced W bosons. 

In \qqlv\ events, where the charge of the W boson decaying leptonically
is known, the likelihood function uses the product of the charge of the lepton 
tagged in the charm decay and the charge of the lepton from the 
leptonically decaying W.
Finally, for electron candidates, the likelihood discriminant employs 
the output of the conversion finder ANN to suppress electrons 
from photon conversions.

In Figure~\ref{fig:tagcontrol}b the likelihood output of the lepton tag 
is shown for data and Monte Carlo at $\sqrt{s} = 183-189$\,GeV. 
In this analysis precise modelling of the lepton identification is 
important. The lepton reconstruction efficiencies in the Monte Carlo
simulation were tuned to match the values obtained from data collected 
at the \Zz\ resonance.
Figures~\ref{fig:syst}b-c show the $E/p$ and normalised $\der E/\der x$
distributions of electron candidates. Reasonable agreement between 
data and Monte Carlo simulation is observed. The variables used for 
muon identification are also well described in the simulation.

%%%%%%%%%%%%%%%%%%%%%%%%%%%%%%% Combined tag %%%%%%%%%%%%%%%%%%%%%%%%%%%%%

\subsection{Combined Tag}
\label{combinedtag}  

For each jet, the vertex ANN and the lepton likelihood
discriminant are merged into a single quantity by an additional 
likelihood function. 
This combined likelihood discriminant uses
two additional variables:
the cosine of the jet angle $\cos \theta_{\mathrm{jet}}$
in the rest frame of the W boson with respect to the W momentum 
direction and the likelihood output of the \WW\ event 
selection~\cite{bib:PR260, bib:PN378}. The jet angle in the di-jet rest 
frame helps to separate jets originating from
up--type and down--type quarks.
Due to the polarisation of the \Wboson\ bosons and the
$V-A$ nature of the weak interaction, up--type quarks
tend to be produced preferentially in the backward direction 
with respect to the \W\ flight direction,
while down--type quarks are found more often in the forward direction. 
The likelihood output of the \WW\ event selection suppresses false tags from
background events, especially from \Zbb\ events. 

%----------------------- TABLE VALUES ---------------------------
\begin{table}[t]
\begin{center}
\begin{tabular}{|c||c|c|c|c|}
\hline
\multicolumn{1}{|c||}{Tagging} & \multicolumn{2}{|c|}{183\,GeV} & \multicolumn{2}{|c|}{189\,GeV} 
\\ \cline{2-5}
 Efficiency (\%)  & \qqlv\ & \qqqq\  & \qqlv\ & \qqqq\ \\ 
  \hline
   $\epsu\ $ & $~4.3\pm0.5$ & $~3.8\pm0.3$ & $~4.9\pm0.5$ & $~4.4\pm0.3$ \\
   $\epsd\ $ & $~3.1\pm0.4$ & $~3.0\pm0.2$ & $~3.2\pm0.4$ & $~3.1\pm0.2$ \\
   $\epss\ $ & $~3.2\pm0.4$ & $~3.0\pm0.2$ & $~3.3\pm0.4$ & $~3.1\pm0.2$ \\
   $\epsc\ $ & $18.1\pm0.6$ & $15.8\pm0.4$ & $18.9\pm0.6$ & $16.7\pm0.4$ \\
\hline
 $\epsbgd\ $ & $~6.0\pm2.7$ & $~4.7\pm1.0$ & $~8.0\pm2.1$ & $~5.4\pm0.8$ \\
\hline
\end{tabular}
\end{center}
\caption[]{\label{tab:tageffs}
  Tagging efficiencies (in \%) after the \WW\ selection from 
  Monte Carlo simulation for u, d, s, and c jets 
  in \WW\ events and for jets from background events, for a cut on
  the combined likelihood of 0.7.
  The values are shown separately for events at
  183\,GeV and 189\,GeV and for \qqlv\ and \qqqq\ events.
  The errors are statistical only.
 }
\end{table} 
%-----------------------------------------------------------------------

\begin{figure}
\unitlength1cm
  \epsfxsize=\textwidth
  \epsfbox{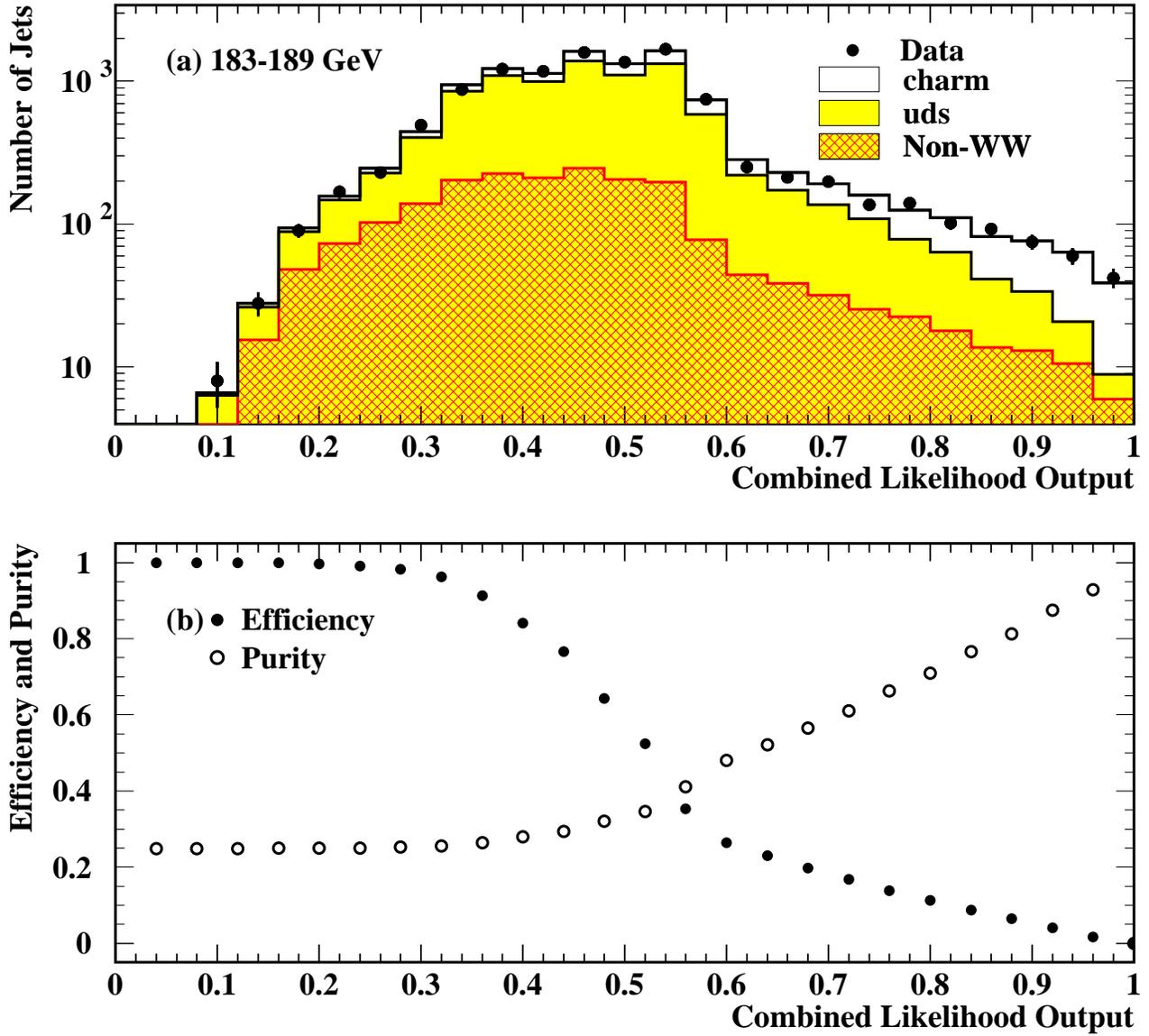}
\caption[]
        {\label{fig:comblike}
        (a) Output of the combined likelihood used to tag charm hadrons
        at $183-189$\,GeV. Points with error bars represent the 
        data (183\,GeV and 189\,GeV samples combined). Histograms
        represent the results of the fits described in Section~\ref{Rcfit}. 
        (b) Efficiency and purity computed after the \WW\ selection as a 
        function of a cut on the combined likelihood output.}
\end{figure}

In Figure~\ref{fig:comblike}a the combined likelihood output
is shown,
illustrating the discriminating power between charm and light 
flavoured jets.
The efficiency and purity for a particular cut on the
combined likelihood output are also depicted in Figure~\ref{fig:comblike}b.
Table \ref{tab:tageffs} lists the tagging efficiencies for jets
in \WW\ and background events that result from a cut on the likelihood 
output at the optimal value of 0.7. 
The tagging efficiencies at 189\,GeV are slightly higher due to 
the extra boost of the \Wboson\ bosons. 
These values are not directly used to derive \Rc, but 
are quoted to give an impression of the performance of the 
charm tag.

\subsection{\boldmath Determination of \Rc}
\label{Rcfit} 

To extract \Rc\ from the data, a binned maximum likelihood fit is 
performed to the combined likelihood distributions. Template histograms
for the charm, light--flavoured, and non--\WW\ background components
are produced from 
the main Monte Carlo samples. 
They are referred to as the probability density 
functions (pdf) for the various components.
Using this fitting method 
improves the statistical sensitivity of the measurement
compared to simply counting the number of jets that pass
a cut on the likelihood output.
The function used to fit the data (see Figure~\ref{fig:comblike}a) is:
\begin{eqnarray}
-\ln {\cal L} &\!=& \!-\sum_{i=1}^N n_i \, 
      \ln \left \{ (1-F_{{\rm{bkg}}})\left (\half \Rc\, {\cal P}_c^i+[1-\half \Rc] \,
      {\cal P}_{\uds}^i \right )
       \!+ F_{{\rm{bkg}}} \, {\cal P}_{{\rm{bkg}}}^i\right \},
\end{eqnarray}
with
\begin{description}
\item[\hspace{0.5in}$i$:] Bin index in the combined likelihood distribution;
\item[\hspace{0.5in}$n_i$:] Number of jets in the $i^{\rm th}$ bin;
\item[\hspace{0.5in}$N$:] Number of bins in the combined likelihood distribution;
\item[\hspace{0.5in}${\cal P}_c^i$:] Probability for a charm jet to appear in the 
                    $i^{\rm th}$ bin (\Wcx\ pdf);
\item[\hspace{0.5in}${\cal P}_{\uds}^i$:] Probability for a \uds\ jet to appear in the
                    $i^{\rm th}$ bin (\Wux\ pdf);
\item[\hspace{0.5in}$F_{{\rm{bkg}}}$:] Fraction of background jets;
\item[\hspace{0.5in}${\cal P}_{{\rm{bkg}}}^i$:] Probability for a  
non--\WW\ jet to appear in the $i^{\rm th}$ 
bin (non--\WW\ background pdf).
\end{description}
The result of the fit for \Rc\ is:
$$
  \Rc = \RcvalueI \pm \RcstatI {\mbox{~~~~at 183~GeV}}
$$
and
$$
  \Rc = \RcvalueII \pm \RcstatII {\mbox{~~~~at 189~GeV}}.
$$
where the errors are statistical only. The $\chi^2$/(degree of freedom) of the fits are
17.2/24 at 183~GeV and 35.4/24 at 189~GeV.

%%%%%%%%%%%%%%%%%%%%%%%%%%%%%%% Systematics %%%%%%%%%%%%%%%%%%%%%%

\subsection{Systematic uncertainties}
\label{life:systematics} 

\begin{figure}
\unitlength1cm
  \epsfxsize=\textwidth
  \epsfbox{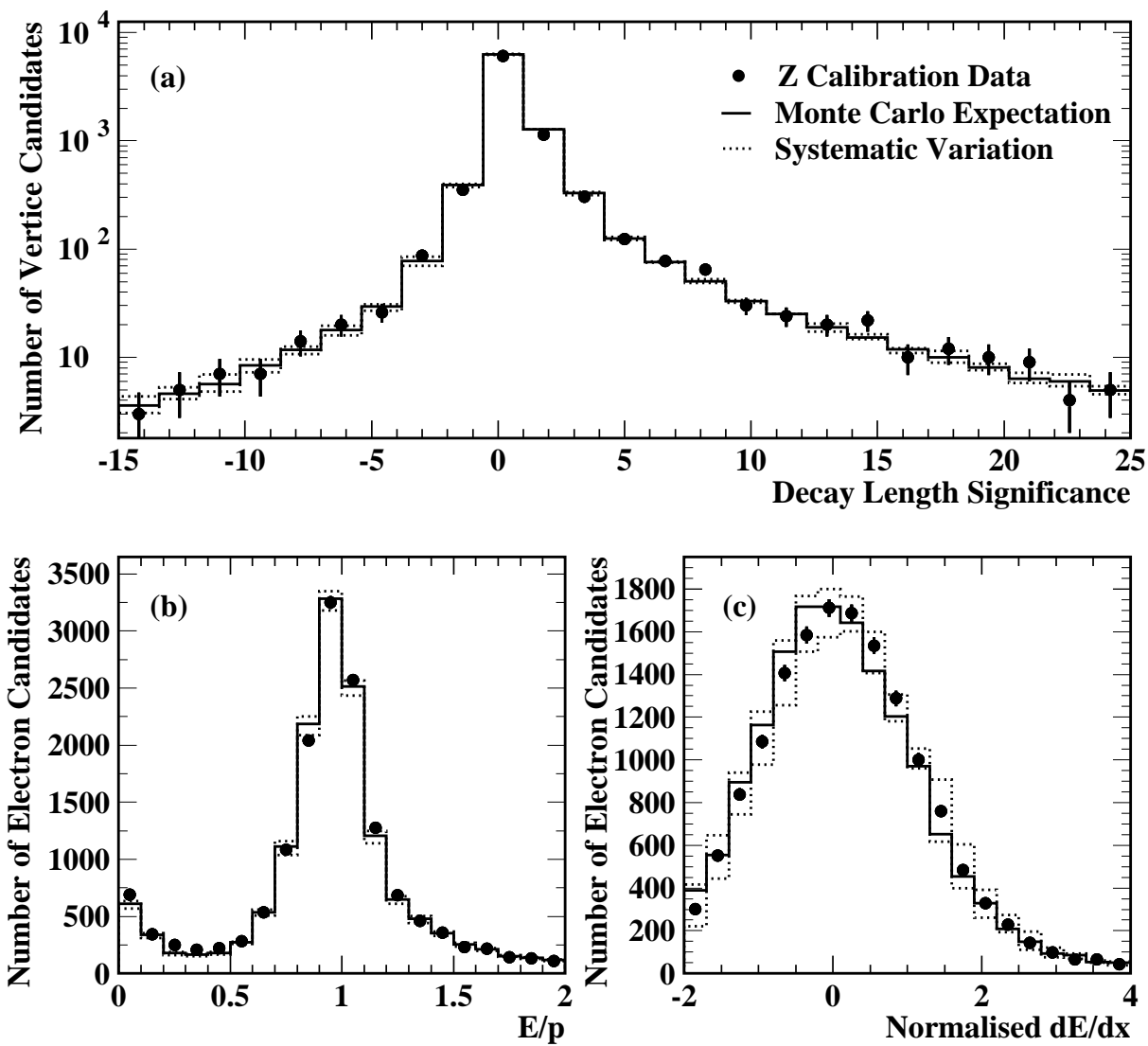}
\caption[]
        {\label{fig:syst}
         Comparison of discriminating variables between calibration data
         taken at the \Zz\ resonance and Monte Carlo prediction.
         Points with error bars represent data from the \Zz\ calibration 
         runs. Histograms show the expectation from the Monte Carole simulation.
         (a) Decay length significance $L/\sigma_L$ for data and Monte Carlo.
         The dotted line shows the effect of the $\pm$5\,\% variation
         of the tracking resolution used to assess the
         systematic uncertainty on the vertex tag.
         (b) $E/p$ and (c) normalised $\der E/\der x$
         for electron candidates from data and Monte Carlo. 
         In these plots, electrons must satisfy the pre-selection 
         described in the text and have an electron ANN output greater 
         than 0.5. The dotted lines show how the variation of the 
         detection (mis)identification and resolution for electron 
         candidates affects the $E/p$ and $\der E/\der x$ distributions.}
\end{figure}

Since the reference histograms, the efficiencies, and
the background for the calculation of \Rc\
were taken from a Monte Carlo simulation, 
the analysis depends on the proper 
modelling of the data distributions in the simulation. 
The sources of systematic error are listed 
in Table~\ref{tab:syserr} and are discussed below. 
Unless otherwise noted, all changes were applied to signal and
background Monte Carlo samples.
\begin{itemize}
\item \emph{Hadronisation model:} 
  The \Pythia\ Monte Carlo sample, which uses the Lund string
  fragmentation as implemented in \Jetset \cite{bib:JETSET}, 
  was compared to a sample generated with \Herwig \cite{bib:HERWIG}, which 
  uses cluster fragmentation. The samples differ in their parton showering
  and hadronisation.
\item \emph{\W\ mass and centre--of--mass energy:}
  As the jet pairing depends on the \W\ mass, \mW,
  and the centre--of--mass energy, $\sqrt{s}$, the analysis 
  was repeated with different 
  \W\ masses and at different centre--of--mass energies. The \W\ mass 
  was varied by 100\,MeV, which covers both the experimental
  uncertainty and the differences between the 
  current world average mass~\cite{bib:PDG} and the value used in 
  the Monte Carlo.
  The centre--of--mass energy was varied by the difference between the 
  luminosity--weighted centre--of--mass energies in data
  and the value of $\sqrt{s}$ used in the main Monte Carlo samples.
\item \emph{Charm fragmentation:} 
  At $\sqrt{s}=M_{\Zboson}$, the mean scaled energy of weakly 
  decaying charm hadrons was found to be 
  $\langle x_{\rm D}(M_{\Zboson})\rangle=0.484\pm0.008$~\cite{bib:LEPEWWG}.
  No measurements of this quantity exist for \Wboson\ decays, 
  but it is expected to be similar to that in \Zboson\ decays. 
  The changes in the charm fragmentation due to QCD effects were taken 
  from \Jetset \cite{bib:JETSET}. To assess the 
  systematic uncertainty, $\langle x_{\rm D}(M_{\Wboson})\rangle$ was varied 
  by the experimental uncertainty obtained 
  on $\langle x_{\rm D}(M_{\Zboson})\rangle$, \ie~$0.008$.
  The fragmentation schemes of Peterson~\cite{bib:Peterson}, Collins and 
  Spiller~\cite{bib:fcolspil}, Kartvelishvili~\cite{bib:fkart} and the Lund 
  group~\cite{bib:flund} were used to assess a possible discrepancy on the
  shape of the fragmentation function.
  The largest variation was assigned as the systematic error.
\item \emph{Background cross--section:}
      In the fit for extracting \Rc, the fraction of background
      was varied according to
      the error on the expected fraction of background events 
      from References~\cite{bib:PR260,bib:PN378}.
\item \emph{Background composition:}
      The various background sources have different probabilities to
      fake a charm jet. 
      The background composition was varied within the 
      uncertainties given in Table~\ref{tab:NW189}.
\item \emph{Charm hadron fractions:} 
  The fractions of the weakly decaying charm hadrons 
  were varied within their experimental errors according to
  Reference~\cite{bib:LEPEWWG}. 
\item \emph{Light quark composition:}
      The shape of the reference
      histogram for light--flavoured quarks depends on the 
      relative fractions of the \uquark, \dquark, 
      and \squark\ jets. 
      The quark flavour fractions were varied 
      according to the actual errors on the CKM 
      matrix~\cite{bib:PDG} so that the error
      covers the effect of assumptions concerning the CKM matrix 
      elements on the \uds\ template histogram. 
      The number of bottom hadrons from \Wboson\ bosons was also 
      doubled to ensure that the measured \Rc\ was insensitive to the 
      bottom fraction in \Wboson\ decays.
\item \emph{Vertex reconstruction:} 
  The correct modelling of the detector resolution in the Monte
  Carlo simulation is of paramount importance for the description of 
  the probability to find a secondary vertex.
  This was checked by comparing the track parameters
  between calibration data collected at $\sqrt{s} \sim M_{\Zboson}$ 
  and Monte Carlo simulation.
  The sensitivity to the vertex reconstruction
  was assessed by degrading or improving the tracking resolution 
  in the Monte Carlo. It was found that changing the track parameters 
  resolution by $\pm5\%$
  covers the range of the observed differences between data and Monte 
  Carlo (see Figure~\ref{fig:syst}a) in the vertex ANN input and output 
  distributions. 
\item \emph{Charm hadron lifetimes:} 
  The lifetimes of the weakly decaying charm hadrons 
  were varied within their experimental errors according to
  Reference~\cite{bib:PDG}. 
\item \emph{Charm decay multiplicity:} 
  The performance of the secondary vertex finder depends on the multiplicity 
  of the final state of the charm hadrons. The relative 
  abundance of 1, 2, 3, 4, 5, and 6 prong charm decays was 
  changed in the Monte Carlo within their experimental
  errors~\cite{bib:markIII}. Similarly, the relative abundance of
  neutral pions was varied in the Monte Carlo within the 
  experimental errors~\cite{bib:markIII}. 
\item \emph{Lepton identification:}
      The modelling of the input variables of the ANN for electron 
      identification 
      has been intensively studied at LEP1~\cite{bib:NN8}, while the 
      performance of the muon tagging ANN has been investigated using various 
      control samples of calibration data collected at the \Zz\ resonance 
      during each 
      year of LEP2~\cite{bib:muNN}. The relative error of the efficiency 
      to identify a genuine 
      electron (muon) has been estimated to be 4\,\% (3\,\%), and
      the relative error on the probability to wrongly identify a
      hadron as a lepton has been evaluated to be around 16\,\% (10\,\%) for 
      electrons (muons)~\cite{bib:NN8, bib:muNN}. 
      By studying the $\der E/\der x$ distributions for electron 
      candidates in calibration 
      data taken at the \Zz\ mass in 1997 and 1998, it was found that
      a shift of the mean $\der E/\der x$ of 0.04~keV/cm (0.3\%) and a scaling
      by 1\% of the $\der E/\der x$ resolution in data was needed to
      correct a slight discrepancy between Monte Carlo and data.
      This effect was also taken into account in the 
      assignment of the uncertainty on the electron tagging efficiency. 
      The (mis)identification probabilities for electrons and muons
      were varied by reweighting the Monte Carlo events
      and the largest resulting change in \Rc\
      was taken as the systematic error. It was verified that 
      the variation of efficiencies and $\der E/\der x$ resolution was sufficient 
      to describe the possible differences between data and Monte 
      Carlo (see Figures~\ref{fig:syst}b-c).       
\item \emph{Lepton energy spectrum from semileptonic charm decays:}
      The energy spectrum of the lepton in the rest frame of the 
      weakly decaying charm hadron
      was reweighted from the spectrum implemented in
      \Jetset\ to the ISGW~\cite{bib:ISGW} and ACCMM~\cite{bib:ACCMM} models;
      the parameters $p\sub{f}$ and $m\sub{s}$ of the ACCMM model were
      varied in the range recommended in Reference~\cite{bib:LEPEWWG}.
\item \emph{Branching ratio \BRctol:}
      The branching fraction for inclusive charm decays to leptons 
      was varied within the range 
      $\BRctol = 0.098 \pm 0.005$~\cite{bib:LEPEWWG} by reweighting 
      the Monte Carlo events. 
\end{itemize}
The size of the systematic errors is very similar for the samples at
183\,GeV and 189\,GeV.
The dominant systematic errors are those associated with the
vertex reconstruction and lepton identification.

%-------------------------- TABLE SYSTEMATICS --------------------------
\begin{table}[t]
\begin{center}
\begin{tabular}{|c||c|c|}
\hline
Source of                    &  \multicolumn{2}{|c|} {$\Delta$ \Rc} \\
Systematic Error             &  183 GeV         & 189 GeV                 \\
\hline\hline
Hadronisation Model            & \EhadrI    & \EhadrII     \\
Centre--of--mass Energy        & \EsqrtsI   & \EsqrtsII    \\
Mass of the \Wboson\ Boson     & \EmwI      & \EmwII       \\
Charm Fragmentation Function   & \EfragI    & \EfragII     \\
Background Cross-Section       & \EbgxsecI  & \EbgxsecII   \\
Background Composition         & \EbgcompI  & \EbgcompII   \\
Charm Hadron Fractions         & \EcfracI   & \EcfracII    \\
Light Quark Composition        & \EudscompI & \EudscompII  \\
Vertex Reconstruction          & \EvtxrecoI & \EvtxrecoII  \\
Charm Hadron Lifetimes         & \EclifeI   & \EclifeII    \\
Charm Decay Multiplicity       & \EcmultI   & \EcmultII    \\
Lepton Identification          & \ElepidI   & \ElepidII    \\
Lepton Energy Spectrum         & \ElepmomI  & \ElepmomII   \\
Branching ratio \BRctol        & \EbrctolI  & \EbrctolII   \\
\hline
Total systematic error         & \RcsystI   & \RcsystII    \\ 
\hline\hline
Statistical error              & \RcstatI   & \RcstatII    \\
\hline\hline
Value of \Rc                   & \RcvalueI  & \RcvalueII   \\
\hline
\end{tabular}
\end{center}
\caption[]{\label{tab:syserr}
  Summary of the experimental systematic errors, statistical errors, and
  central values from the binned maximum likelihood fit on \Rc\ at 183\,GeV 
  and 189\,GeV.
}
\end{table} 
%-------------------------------------------------------------------

\section{\boldmath Results}
\label{Rcwresult} 

The result of the charm tag analysis is 
$$
  \Rc = \RcvalueI \pm \RcstatI \stat \pm \RcsystI \syst {\mbox{~~~~at 183~GeV}}
$$
and
$$
  \Rc = \RcvalueII \pm \RcstatII \stat \pm \RcsystII \syst {\mbox{~~~~at 189~GeV}}.
$$
The combined value of \Rc obtained at $\sqrt{s}=183$\,GeV 
and 189\,GeV is
$$
  \Rc = \Rcvalue \pm \Rcstat \stat \pm \Rcsyst \syst,
$$
where the systematic uncertainties are treated as being fully correlated.

%%%%%%%%%%%%%%%%%%%%%%%%%%%%%%% Extraction of Vcs %%%%%%%%%%%%%%%%%%%%%%

\section{\boldmath Extraction of \Vcs}
\label{Vcsextraction} 

%------------------------- TABLE VCS VALUES ---------------------------
\begin{table}[t]
\begin{center}
\begin{tabular}{|c|l@{$\,\pm\,$}l|}
\hline
CKM matrix element & \multicolumn{2}{|c|}{Value}   \\ 
\hline\hline
  \Vud      & 0.9735 & 0.0008 \\
  \Vus      & 0.2196 &  0.0023 \\
  \Vub/\Vcb & 0.090~ &  0.025~ \\
  \Vcd      & 0.224~ &  0.016~ \\
  \Vcb      & 0.0402 &  0.0019 \\
\hline
\end{tabular}
\end{center}
\caption[]{\label{tab:Vcsvalues}
  Values of the CKM matrix elements used in the calculation of \Vcs\
  from \Rc. The values are taken from Reference~\cite{bib:PDG}.
%% Note: Values are updated from 1999 web update: 
%% http://www-pdg.lbl.gov/1999/kmmixrpp.ps
%% Additional note: The values have NOT changed for the Y2K update
  }
\end{table}
%-----------------------------------------------------------------------

It is possible to extract \Vcs\ from \Rc\ using the relation
\begin{equation}
  \Rc = \frac{\Vcd^2 + \Vcs^2 + \Vcb^2}
             {\Vud^2 + \Vus^2 + \Vub^2 + \Vcd^2 + \Vcs^2 + \Vcb^2},
\end{equation}
which leads to 
\begin{eqnarray}
 \Vcs  & = & \sqrt{\frac{\Rc}{1-\Rc} \left( \Vud^2 + \Vus^2 + \Vub^2 \right)
                   - \Vcd^2 - \Vcb^2}.
 \label{eq:Vcs}
\end{eqnarray}
Using the values of the other CKM matrix elements listed in 
Table~\ref{tab:Vcsvalues}, one gets
$$
  \Vcs = \Vcsfinal \pm \Vcsckm\,\ckmerr,
$$
where the last error is due to the uncertainty on the CKM matrix
elements used in the calculation. 

A more precise value for \Vcs\ can be obtained based on
the Standard Model prediction of the decay width of the \Wboson\ boson to
final states containing charm quarks:
\begin{equation}
  \Gamma\,(\W \rightarrow \mathrm{c\,X}) = 
                     \frac{C \GF \Mw^3}{6 \sqrt{2} \pi}\, 
                     (\Vcd^2 + \Vcs^2 + \Vcb^2),
%                     \approx 
%                     (\Vcd^2 + \Vcs^2 + \Vcb^2)
%                      \, (705 \pm 4)\,\MeV,
%% Updated 707+-1 -> 705+-4, 19.7.00
\label{eq:gammaVcs}
\end{equation}
with $\frac{C \GF \Mw^3}{6 \sqrt{2} \pi}=(705 \pm 4)\,\MeV$~\cite{bib:PDG} 
and the color factor $C$ given by~\cite{bib:PDG}
%% Note: Has been updated in the Y2K review: 707+-1 -> 705+-4
%% BL, 19.7.00
\begin{equation}
  C =  3 \left ( 1 + \frac {\alphas\,(\Mw)}{\pi} + 
                            1.409 \frac {\alphas^2\,(\Mw)}{\pi^2}  
                            -12.77 \frac {\alphas^3\,(\Mw)}{\pi ^3} \right ) .
\end{equation}
$\Gamma\,(\W \rightarrow \mathrm{c\,X})$
can be evaluated from \Rc\ using the relation
\begin{equation}
  \Gamma\,(\W \rightarrow \mathrm{c\,X}) = \Rc \,
    \Br\,(\Wtohad) \, \GammatotW.
\label{eq:gammaRc}
\end{equation}

Using the {\rm{PDG}} values $\Br\,(\Wtohad) = 0.6848 \pm 0.0059$ 
%% Updated 19.7.00 BL to PDG 2000 value
and\footnote{
  The most precise measurements of $\GammatotW$ are determined
  from the quantity 
  $\BigR = \frac{\sigma\,({\mathrm{\ppbar \to \W + X}}) \cdot \Br\,(\Wtoev)} 
                {\sigma\,({\mathrm{\ppbar \to \Zz + X}}) \cdot \Br\,(\Ztoee)}$.
  In~\cite{bib:d0cdf}, the coupling of the \Wboson\ boson to light quarks is 
  needed as 
  input; however, direct measurements of the corresponding CKM matrix 
  elements are so
  precise that it introduces a negligible uncertainty on $\GammatotW$.
}
$\GammatotW = (2.12 \pm 0.05)\,\GeV$ \cite{bib:PDG}
%% Has been updated for Review 2000, BL 19.7.00
together with the measurement of \Rc\ presented in this letter,
one gets
$$
  \Gamma\,(\W \rightarrow \mathrm{c\,X}) = \GammaWfinal,
$$
where the last error is due to the uncertainties on the external 
parameters \GammatotW\ and $\Br\,(\Wtohad)$. This corresponds to
$$
  \Vcd^2 + \Vcs^2 + \Vcb^2 = \SumVijfinal.
$$
Subtracting the off--diagonal CKM matrix elements, using the values listed 
in Table~\ref{tab:Vcsvalues}, results in
$$
 \Vcs = \VcsfinalG,
$$
where the last error is due 
to the uncertainties on the CKM matrix elements used in the 
calculation. 
The determination of \Vcs\ from Equations~(\ref{eq:gammaVcs}) 
and~(\ref{eq:gammaRc}) 
only involves the CKM matrix elements for the transition 
$\W \rightarrow \mathrm{c\,X}$ and
well measured parameters such as \GammatotW\ and $\Br\,(\Wtohad)$. 
It is therefore more precise than the method which relies on
Equation~(\ref{eq:Vcs}).
Both results of \Vcs\ presented here are 
compatible with the direct determination from $D$ semielectronic decays 
which yields $\Vcs=1.04 \pm 0.16$~\cite{bib:PDG}.

%%%%%%%%%%%%%%%%%%%%%%%%%%%%%%% Summary %%%%%%%%%%%%%%%%%%%%%%%%%%%%%%%%

\section{Summary}
\label{summary} 

Using data taken with the OPAL detector at LEP at
centre--of--mass energies of 183\,GeV and 189\,GeV,
the ratio 
 $\Rc = \Gamma\,(\W \rightarrow \mathrm{c\,X}) / 
    \Gamma\,(\W \rightarrow \mathrm{hadrons})$
has been measured using a technique based on jet properties, lifetime 
information, and leptons from charm decays. The combined result is 
$$
  \Rc = \Rcfinal,
$$
in agreement with the Standard Model expectation of 0.5.
This result is consistent with the Standard Model prediction
that the \Wboson\ boson couples to up and charm quarks with equal 
strength and it is in agreement with the recent measurements 
of DELPHI~\cite{bib:rcwdelphi} and ALEPH~\cite{bib:rcwaleph}. 

Using the results from direct measurements of the other CKM matrix
elements, \Vcs\ can be calculated from \Rc, yielding $\Vcs = \Vcssummary$.
A more precise value for \Vcs\ can be derived from
the measurements of \Rc, $\Br\,(\Wtohad)$, and $\GammatotW$:
$$
 \Vcs = \VcssummaryG.
$$
This can be compared to a more indirect determination of \Vcs\ by OPAL 
from the ratio of the \Wboson\ decay widths to \qq\ and \lv, which 
yields $ \Vcs = 1.014 \pm 0.029 \stat \pm 0.014 \syst$~\cite{bib:PN378} 
%% Updated AB 20.7.00 from final draft of W cross section paper 
under the assumption that the \Wboson\ boson couples with equal strength to 
quarks and leptons.
It should be noted that the ratio $\Br\,(\Wqq)/\Br\,(\Wlv)$ is not used
in the analysis presented here, 
so that these are independent determinations of \Vcs.

%%%%%%%%%%%%%%%%%%%%%%%%% Acknowledgements %%%%%%%%%%%%%%%%%%%%%%%
\section*{Acknowledgements}

We particularly wish to thank the SL Division for the efficient
operation
of the LEP accelerator at all energies
 and for their continuing close cooperation with
our experimental group.  We thank our colleagues from CEA, DAPNIA/SPP,
CE-Saclay for their efforts over the years on the time-of-flight and
trigger
systems which we continue to use.  In addition to the support staff at
our own
institutions we are pleased to acknowledge the  \\
Department of Energy, USA, \\
National Science Foundation, USA, \\
Particle Physics and Astronomy Research Council, UK, \\
Natural Sciences and Engineering Research Council, Canada, \\
Israel Science Foundation, administered by the Israel
Academy of Science and Humanities, \\
Minerva Gesellschaft, \\
Benoziyo Center for High Energy Physics,\\
Japanese Ministry of Education, Science and Culture (the
Monbusho) and a grant under the Monbusho International
Science Research Program,\\
Japanese Society for the Promotion of Science (JSPS),\\
German Israeli Bi-national Science Foundation (GIF), \\
Bundesministerium f\"ur Bildung und Forschung, Germany, \\
National Research Council of Canada, \\
Research Corporation, USA,\\
Hungarian Foundation for Scientific Research, OTKA T-029328, 
T023793 and OTKA F-023259.\\

%%%%%%%%%%%%%%%%%%%%%%%%% References %%%%%%%%%%%%%%%%%%%%%%%%%%%%%

\clearpage

\clearpage 
%%%%%%%%%%%%%%%%%%%%%%%%%%%%%%%% FIGURES %%%%%%%%%%%%%%%%%%%%%%%%%%%%%%%

\begin{thebibliography}{99}

\bibitem{bib:CKMmatrix}
N.~Cabibbo, \PRL{10}{1963}{531}; \\
M.~Kobayashi and T.~Maskawa, Prog.\ Theor.\ Phys.\ \textbf{49} (1973) 652.

\bibitem{bib:PDG}
D.E..~Groom \etal, \EPC{15}{2000}{1}.

\bibitem{bib:OPALdetector}
 OPAL Collaboration, K.~Ahmet \etal, \NIMA{305}{1991}{275}; \\
 P.P.~Allport \etal, \NIMA{324}{1993}{34};                  \\
 P.P.~Allport \etal, \NIMA{346}{1994}{476}; \\
 S.~Anderson \etal, \NIMA{403}{1998}{326}.

\bibitem{bib:KORALW} M.~Skrzypek \etal, \CPC{94}{1996}{216}; \\
 M.~Skrzypek \etal, \PLB{372}{1996}{289}.

\bibitem{bib:JETSET}
 T. Sj\"ostrand, \CPC{39}{1986}{347}; \\  
 T. Sj\"ostrand and H.--U. Bengtsson, \CPC{43}{1987}{367}.

\bibitem{bib:GOPAL} J.~Allison \etal, \NIMA{317}{1992}{47}.
  
\bibitem{bib:PYTHIA}
 T.~Sj\"ostrand, \CPC{82}{1994}{74}.  
  
\bibitem{bib:HERWIG} G.~Marchesini \etal, \CPC{67}{1992}{465}.

\bibitem{bib:Excalibur} F.A.~Berends, R.~Pittau and R.~Kleiss, 
\CPC{85}{1995}{437}.

\bibitem{bib:GRC4F} J.~Fujimoto \etal, \CPC{100}{1997}{128}.

\bibitem{bib:PR260} 
{\opalabbiendi, \EPC{8}{1999}{191}}.

\bibitem{bib:PN378} OPAL Collaboration,
  {\itshape \WpWm\ Production Cross Section and \Wboson\ 
  Branching Fractions in \epem\ Collisions at 189\,{\rm GeV},}
  OPAL PR321, CERN-EP-2000-101 (2000), Submitted to Phys. Lett. B.

\bibitem{bib:Durham} S.~Catani \etal, \PLB{269}{1991}{432}.

\bibitem{bib:Wmass183} \opalabbiendi, \PLB{453}{1999}{138}.

\bibitem{bib:Wmass189} OPAL Collaboration,
  {\itshape Measurement of the Mass and Width of the \Wboson\ Boson in
  \epem\ Collisions at 189,{\rm GeV},}
  OPAL PR320, CERN-EP-2000-099 (2000), Submitted to Phys. Lett. B.

\bibitem{bib:PN342}
{\opalabbiendi, \PLB{453}{1999}{153}}.

\bibitem{bib:teardown}
\opalackerstaff, \ZPC{74}{1997}{1}.

\bibitem{bib:OPALRc} 
{\opalackerstaff, \EPC{1}{1998}{439}}.

\bibitem{bib:Jetnet}
  C.~Peterson, T.~R\"ognvaldsson and L.~L\"onnblad, 
  {\itshape JETNET 3.0 - A Versatile Artificial Neural Network Package,}
  LU--TP--93--29, CERN--TH.7135/94.

\bibitem{bib:joint}
  ALEPH Collaboration, D.~Buskulic \etal,
  \PLB{313}{1993}{535}.

\bibitem{bib:NN5}
\opalalexander, \ZPC{70}{1996}{357}.

\bibitem{bib:NN8}
{\opalabbiendi, \EPC{8}{1999}{217}}.

\bibitem{bib:muNN}
     OPAL Collaboration,
     {\itshape Measurements of $R_{\mathrm{b}}$, $A^{\mathrm{b}}_{\mathrm{FB}}$, and 
       $A^{\mathrm{c}}_{\mathrm{FB}}$ in ${\mathrm{e^+e^-}}$ Collisions 
       at 130--189\,{\rm GeV},} CERN-EP/99-170 (1999), Submitted to Eur. Phys. J. C.

\bibitem{bib:PCpackage}
  D.~Karlen,
  \CiP{12:4}{1998}{380}.

\bibitem{bib:LEPEWWG}
  ALEPH, DELPHI, L3, and OPAL Collaborations: \NIMA{378}{1996}{101}; \\
  Updates of the recommendations are given in the LEP electroweak working group: 
  {\it Input Parameters for the LEP Electroweak Heavy Flavour Results for Summer
  1998 Conferences}, internal note LEPHF/98--01 (URL: {\tt 
  http://www.cern.ch/LEPEWWG/heavy/}).
 
\bibitem{bib:Peterson}
  C.~Peterson, D.~Schlatter, I.~Schmitt, and P.M.~Zerwas,
  \PRD{27}{1983}{105}.

\bibitem{bib:fcolspil}
P.~Collins and T.~Spiller, \JPH{G11}{1985}{1289}.

\bibitem{bib:fkart}
V.G.~Kartvelishvili, A.K.~Likhoded and V.A.~Petrov, \PLB{78}{1978}{615}.

\bibitem{bib:flund}
B.~Anderson, G.~Gustafson and B.~S\"oderberg, \ZPC{20}{1983}{317}.

\bibitem{bib:markIII}
Mark III Collaboration, D.~Coffman \etal, \PLB{263}{1991}{135}.
 
\bibitem{bib:ISGW}
N.~Isgur, D.~Scora, B.~Grinstein, M.B.~Wise, \PRD{39}{1989}{799}; \\
N.~Isgur and M.B.~Wise, \PRD{41}{1990}{151}.

\bibitem{bib:ACCMM}
  G.~Altarelli \etal,
  \NPB{208}{1982}{365}.

\bibitem{bib:d0cdf}
D0 Collaboration, S.~Abachi \etal, \PRL{75}{1995}{1456}; \\
CDF Collaboration, F.~Abe \etal, \PRD{52}{1995}{2624}.

\bibitem{bib:rcwdelphi}
DELPHI Collaboration, P.~Abreu \etal, 
\PLB {439}{1998}{209}.

\bibitem{bib:rcwaleph}
ALEPH Collaboration, R.~Barate \etal, 
\PLB {465}{1999}{349}. 
\end{thebibliography}
\end{document}